\documentclass[12pt]{amsart}

\usepackage{amssymb,amscd}\usepackage{pb-diagram}\usepackage{verbatim}\usepackage{graphicx}

\newtheorem*{Whitney towers}{Theorem~\ref{Whitney towers}}
\newtheorem*{h-towers}{Theorems ~\ref{half} \& \ref{$(n)$-solvable}}

\newtheorem*{surgery curves}{Theorem~\ref{surgery curves}}
\newtheorem*{cg=0}{Theorem~\ref{vanish}}

\theoremstyle{definition}

\numberwithin{equation}{section}
\numberwithin{figure}{section}

\newcommand{\x}{\times}

\newcommand{\N}{\mathbb{N}}

\newcommand{\Q}{\mathbb{Q}}

\newcommand{\f}{\noindent}

\renewcommand{\SS}{\mathcal{S}}

\title{
Remarks on $n$-dimensional Feynman diagrams, \\
for example, which will appear \\
in M-theory and in F-theory 
}
\author{Eiji Ogasa}
\address{
High Energy Physics Theory Group\\
Department of Physics\\
Ochanomizu University\\ 
Bunkyo-ku\\
Tokyo 112-8610\\ 
JAPAN\\}
\email{ ogasa@phys.ocha.ac.jp}

\begin{document}\begin{abstract} 
We state some remarks on `$n$-dimensional Feynman diagrams' ($n\in\N$). 
\end{abstract} 


\maketitle

`$n$-dimensional Feynman diagrams' ($n\in\N$) 
will be used in physics in the near future. 
Here, we let 1-dimensional Feynman diagrams mean  
Feynman diagrams in  `usual' QFT (see \cite{Feynman}\cite{PeskinSchroeder}).  
Furthermore, we let 2-dimensional Feynman diagrams mean 
world sheets in `usual' superstring theory (see \cite{GSW}\cite{Polchinski}). 
We introduce `$n$-dimensional Feynman diagrams' as a generalization of 
the 1-, 2-dimensional Feynman diagrams as follows.

F-theory, M-theory  (see 
\cite{HullTownsend} 
\cite{Johnson}
\cite{Schwarz} 
\cite{Townsend}
\cite{Vafa}  
\cite{Witten} )  etc. 
imply that particles are represented by manifolds 
whose dimensions are greater than one. 
Here, consider interaction of these 
particles and use perturbation theory like the 0-,1-dimensional particle case. 
Then we will be able to use manifolds 
whose dimensions are greater than two  
in order to represent the interaction. 
In this paper we call these manifolds 
$n$-dimensional Feynman diagrams 
or 
$n$-dimensional world membranes 
if the manifolds are $n$-dimensional ones. 
We may use not only manifolds 
but also  CW complexes for particles and Feynman diagrams 
(See \cite{DavisKirk} for CW complexes). 

Suppose we will complete the high-dimensional particle theory 
($F$-theory, $M$-theory etc.) 
without using perturbation theory or Feynman diagrams.  
However, the limit of the theories is  `usual' superstring theory 
or  `usual' QFT. 
Therefore, we will be able to consider $n$-dimensional Feynman diagrams. 

Anyway, mathematically we can discuss $n$-dimensional Feynman diagrams.

In this paper 
we state some remarks on `$n$-dimensional Feynman diagrams.'

\vskip3mm

We consider the case where we make two 3-vertex functions 
into a 4-vertex function. 
In this case there is a different feature between in the case of high dimensional Feynman diagrams and in the 1-, 2-dimensinal case.

Let $m$ be any integer greater than two.  
Let $M$  be a compact oriented connected $m$-manifold  with boundary.  
Let $\partial M=A_1\amalg A_2\amalg A_3$, 
where $\amalg$ denotes a disjoint union and 
$A_1 (resp. A_2, A_3)$ is a connected closed oriented manifold.
Let the orientation of $A_i$ be induced from $M$. 
Suppose that there is an orientation reversing diffeomorphism 
$A_i\to A_j$, where we do not assume $i=j$ or $i\neq  j$. 
Let $M'$ be diffeomorphic to $M$. 
Let $\partial M'=A'_1\amalg A'_2\amalg A'_3$ and 
 $A_l=A'_l (l=1,2,3)$. 
 Take $M$ and $M'$. 
Identify $A_i$ and $A'_j$ by an orientation reversing diffeomorphism 
$f:A_i\to A'_j$  
and obtain a compact oriented connected 
$m$-manifold $W_f$ from $M$ and $M'$. 
Let $\mathcal S_M$ be the set whose elements are 
the diffeomorphism classes of such $W_f$

\vskip3mm
\f{\bf Theorem 1.} {\it 
For $m\geqq3$, there is a compact oriented connected $m$-manifold $M$  
such that the above set $\SS_M$ is an infinite set. }

\f{\bf Note.} 
Theorem 1 means that, for $m\geqq3$, 
two 3-vertx functions can make infinitely many kinds of 4-vertex functions 
under some conditions.  
If $m=2$, such $M$ does not exist 
as string theorists and topologists know.
For $m=1$, they can say such does not.
\vskip3mm

\f{\bf Proof.} 
The $m=3$ case: Take a solid torus. Remove two open 3-balls from the solid 
torus, call it $M$. 
Note $\partial M=S^2\amalg S^2\amalg T^2$. 
Let $f:T^2\to T^2$ be a diffeomorphism. 
Let $Z$ be any 3-dimensional Lens space 
(see \cite{Rolfsen} for Lens spaces). 
Take any oriented manifold which is made from $Z$ 
by removing four open 3-balls, call it $Z'$.  
Then $Z'\in\SS$.  
Hence there are countably infinitely many $Z'$, 
using the homology groups of $Z'$. 
Hence Theorem 1 is true in the $m=3$ case.

The $m>3$ case: 
Take $D^2\x T^{m-2}$, where $T^{m-2}$ is an $(m-2)$-dimensional torus. 
Remove two open $m$-balls from   $D^2\x T^{m-2}$, call it $M$. 
Then $\SS$ includes all manifolds which are made from 
all $[($Lens spaces$)\x T^{m-3}]$
by removing four open $m$-balls, call it $Z'$. 
Hence there are countably infinitely many $Z'$, 
using the homology groups of $Z'$. 
Note $\SS$ is an infinite set. 
Hence Theorem 1 is true in the $m>3$ case. 
This completes the proof of Theorem 1. 

\vskip3mm

We consider the case where 
we make 3-vertex functions into an $l$-vertex function ($l\in\N\cup\{0\}$). 
Here, let the number of kinds of 3-vertex functions be finite. 
In this case there is a different feature 
between in the case of high dimensional Feynman diagrams 
and in the 1-, 2-dimensinal case, too.

Let $m$ be any integer greater than two. 
Let $M_i (i=1,...,\mu)$ be 
a compact oriented connected $m$-dimensional manifold with boundary, 
where $\mu\in\N$.
Let $\partial M_{i}=\amalg_{j=1}^{j=\nu_i} M_{ij}$, 
where 
$\nu_i\in\N$,  $\nu_i\geqq3$, 
$M_{ij}$ is a connected closed oriented manifold, 
and 
the orientation of $M_{ij}$ is induced from that of $M_i$.
Let $B=\amalg B_z (z\in\N, z\geqq3)$  be a closed oriented $(m-1)$-manifold.

We define a set $\mathcal W_{\{M_i\},B}$:
An arbitrary element  $\in\mathcal W_{\{M_i\},B}$ 
is a compact connected oriented manifold 
with boundary $B$ with the following properties;  
There are embedded closed $(m-1)$ manifolds 
$Y_1,...,Y_\alpha\subset$ Int$W$, 
where Int$W$ means the interior of $W$ and  $Y_i\cap Y_j=\phi$ for $i\neq j$. 
Let $N(Y_i)=Y_i\x[-1,1]$ be the neighborhood of $Y_i$ in $W$. 
Take $W-$Int$N(Y_i)=W_1\amalg...\amalg W_w$. 
Then each $W_l$ is diffeomorphic to $M_i$ for an $i$.


Let  $\mathcal X_B$ be a set of all compact oriented connected $m$-manifolds with boundaries $B$.

\vskip3mm
\f{\bf Theorem 2.}  {\it 
Let $m$, $M_i$, $B$, $\mathcal W_{\{M_i\},B}$, and $\mathcal X_B$ 
be as above.  
Then, 
for any $B$ and any $M_i$, we have 
 $\mathcal W_{\{M_i\},B}\neq\mathcal X_B$.  
}   

\f{\bf Note.} 
If $m=2$,  for any $B$ there exists a manifold $M$ such that , 
$\mathcal W_{\{M\},B}=\mathcal X_B$, 
as topologists and string theorists know. 
Here, $\{M\}$ is a set which has only one element $M$. 
Theorem 2 means that we may need a new discussion to divide 
complex Feynman diagrams into fundamental parts. 

\vskip3mm

\f{\bf Proof.}  
Let $W\in\mathcal W_{\{M_i\},B}$. 
Then we can divide $W$ into pieces $W_i$, $N(Y_j)$ as above 
and can regard $W=W_1\cup...\cup W_w$. 
Consider the Meyer-Vietoris exact sequence 
(see \cite{DavisKirk}\cite{MilnorStasheff}\cite{Steenrod} 
for the Meyer-Vietoris exact sequence):

\f
$H_j( \amalg_{i,i'}\{W_i \cap W_{i'}\};\Q)
\to H_j(\amalg_{i=1}^{i=w} W_i;\Q)\to H_j(W;\Q)$. 
Here, $\amalg_{i,i'}$ means the disjoint unions of $W_i \cap W_{i'}$ for all $(i,i')$.
Consider 

\f 
$H_1(W;\Q)\to H_0(\amalg_{i,i'}\{W_i\cap W_{i'}\};\Q)\to
H_0(\amalg W_i;\Q)\to H_0(W;\Q)\to0$. 
Note 

\f
$H_0(\amalg W_i;\Q)\cong\Q^w$ and $H_0(W;\Q)\cong\Q$.

Suppose that $\partial W=B$ has $z$ components as above, 
that is, it is a Feynman diagrams with $z$ outlines. 
Hence 
$H_0(\amalg_{i,i'}\{W_i \cap W_{i'}\};\Q)
\cong\Q^\rho$, where $\rho\geqq{\frac{3w-z}{2}}$. 

We suppose Theorem 2 is not true and 
induce the contradictory. 
If Theorem 2 is not true, then $\mathcal W_{\{M_i\},B}$ includes 
any compact oriented connected $m$-manifold 
with boundary 
$B$.  
If $H_1(W;\Q)\cong\Q^l$, 
we have the exact sequence: 

\f
$\Q^l\to\Q^\rho\to\Q^w\to\Q\to0.$   
Hence $l\geqq\rho-w+1$. 
Hence $l\geqq\frac{w-z+2}{2}$. 
Hence $(2l+z-2)\geqq w$.

Let $X$ be a compact manifold. 
Take a handle decomposition of $X$. 
Consider the numbers of handles in the handle decompositions. 
Let $h(X)$ be the minimum of such the numbers.  
Suppose that $M$ is one of the manifolds $M_i$ and 
that $h(M)\geqq h(M_i)$ for any $i$.  
Then we have $w\times h(M)\geqq h(W)$. 
Hence $(2l+z-2)\times h(M)\geqq h(W)$.

For any natural number $N$, 
there are countably infinitely many compact oriented connected 
$m$-manifolds $W$ with boundaries 
such that $\partial W$ has $z$ components, 
that 

\f
$H_1(W;\Q)\cong\Q^l$, 
and that $h(W)\geqq N$. 
Because: There is an $n$-dimensional manifold $P$ such that 
$H_1(P;\Q)\cong\Q^l$. 
There is an $n$-dimensional rational homology sphere $Q$ 
which is not an integral homology sphere. 
Make a connected sum which is made from one copy of $P$ and $q$ copies of $Q$
($q\in\N\cup\{0\}$).

This is the contradiction.  This completes the proof.

\vskip3mm

We consider the case where 
we make 3-vertex functions into an $l$-vertex function ($l\in\N\cup\{0\}$). 
Here, let the number of kinds of 3-vertex functions be infinite. 

Let  $n$ be any integer greater than two. 
Then there is an infinite set  $\mathcal Q$ 
which is a proper subset of the set of all $n$-manifolds 
(i.e. which is not the set of all $n$-manifolds) 
with the following properties: 
Using elements of a finite subset of $\mathcal Q$ 
in the similar manner 
to make $W$ from 
$\{M_i\}$ 
above Theorem 2, 
we can construct any compact oriented $n$-manifold with boundary.  
Furthermore, we can suppose 
the boundary of any element of $\mathcal Q$ 
has one, two or three components. 
We state theorem in the $n=3$ case. 
In the $n>3$ case we have similar theorem. 
The proof is easy, considering properties of handle decompositions.
See \cite{Smale} \cite{Milnor} for handle decompositions.

\vskip3mm
\f{\bf Theorem 3.} {\it 
Let $Q_g$ be a compact oriented 3-manifold 
whose handle decomposition is 
$(F_g\times [0,1])\cup($a 1-handle$)$, 
where  $F_g$ denotes a closed oriented surface with genus $g$. 
Let $Q_{g,h}$ be a compact connected oriented 3-manifold 
whose handle decomposition is 
$(F_g\times [0,1])\cup (F_h\times [0,1])\cup($a 1-handle$)$.@
Note $\partial Q_g=F_g \amalg F_{g+1}$ and 
$\partial Q_{g,h}=F_g \amalg F_h \amalg F_{g+h}$. 
Take a set $\mathcal Q=\{B^3, Q_{g,h}, Q_g\vert g,h\in\N\cup\{0\}\}$, 
where $B^3$ is a 3-ball. 
Let $M$ be an arbitrary compact oriented 3-manifold with boundary.  
Then $M$ is 
made from elements of a finite subset of $\mathcal Q$
in the similar manner 
to make $W$ from 
$\{M_i\}$ 
above Theorem 2. 
} 
\vskip3mm

One way to use $n$-dimensional Feynman diagrams ($n\geqq3$) is 
to restrict what kind of compact oriented $n$-manifolds
to represent Feynman diagrams. 
Indeed, in the $n=1$ case, we restrict what kind of 
`CW-complexes made from 0-cells and 1-cells' 
represent the Feynman diagrams. See \cite{DavisKirk} for CW complexes. 

We might note the following: Suppose that we use only elements of 
$\mathcal R=$ $\{B^3, Q_{g,}, Q_{g,h}\vert$ $ g, h\in A\}$, 
where $A$ is a finite subset of $\N\cup\{0\}$. 
Then any element of $\mathcal R$ includes a submanifold 
which is diffeomorphic to 
 $Q_{k}$ (resp. $Q_{k,l}$)  
for any $k,l\in\N\cup\{0\}$. 
It might not be a good idea to restrict 
what kind of compact oriented 3-manifolds to represent Feynman diagrams.


Although it is one way of saying, 
Witten's Chern-Simons theory (see \cite{WittenQFTJones})
on 3-manifolds $M$ with the gauge group $G$ 
are regarded 
as the theory $M\to \mathcal G$, 
where  $\mathcal G$ is the Lie ring of $G$. 
Recall that $\mathcal G$ is a vector space. 
Note that, in this case, we can regard all compact oriented 3-manifolds 
with boundaries  as 3-dimensional world membranes. 
It might not be a good idea to restrict unnaturally 
what kind of compact oriented 3-manifolds to represent Feynman diagrams. 
For example, we might need an idea that such restriction make a sense in only low-energy case.

\vskip3mm
In the two dimensional case (i.e. `usual' string theory)  
particles are represented  both by closed manifolds (i.e. closed strings) 
and by compact manifolds with boundaries (i.e. open strings). 
In the $n$-dimensional case ($n\geqq3$) particles will be represented 
both by closed manifolds and by compact manifolds with boundaries. 
In this paper we concentrate on the case of closed manifolds.

It might be good to suppose that $n$-dimensional Feynman diagrams are 
complex manifolds, 
symplectic ones, 
K\"ahler ones, toric ones, 
hyperbolic ones, 
or something. 
However, in these cases, there exist underlying smooth manifolds 
(and underlying topological manifolds). 
Hence our theorems in this paper 
are fundamental restrictions to such the theories,  
as Pauli exculsion rule and Coleman-Mandula NO-GO theorem are. 
Because in  our theorems $n$-dimensional Feynman diagrams 
are just smooth manifolds.

Research on $n$-dimensional Feynman diagrams in a time-space 
is connected with that on submanifolds in a manifold. 
Submanifold theory includes $n$-dimensional knot theory as an important field. 
See \cite{CochranOrr}\cite{LevineOrr}\cite{Ogasa}.



\begin{thebibliography}{ABCD}

\bibitem{CochranOrr}  T. D. Cochran and K. E. Orr: 
Not all links are concordant to boundary links 
{\it Ann. of Math.}, 138, 519--554, 1993. 

\bibitem{DavisKirk} 
J. Davis and P. Kirk: Lecture notes in algebraic topology. 
{\it Graduate Studies in Mathematics, 35. 
American Mathematical Society, Providence, RI}, 2001.


\bibitem{Feynman} 
R. P. Feynman:  Space-time approach to non-relativistic quantum mechanics. 
{\it Rev. Modern Physics 20}, (1948). 367--387. 

\bibitem{GSW} M. B. Green, J. H. Schwarz, and E. Witten:  
Superstring theory Vol. 1, 2.
{\it Cambridge, Uk: Univ. Pr. (Cambridge Monographs On Mathematical Physics).} 
1987.


\bibitem{Johnson}
 C. V. Johnson: D-branes. 
{\it Cambridge Monographs on Mathematical Physics. 
Cambridge University Press, Cambridge}, 2003.



\bibitem{HullTownsend}
C. M. Hull and P. K. Townsend: 
Unity of superstring dualities. 
{\it Nuclear Phys. B 438 (1995), no. 1-2, 109--137}. 



\bibitem{LevineOrr} 
J. Levine and K. Orr: 
A survey of applications of surgery to knot and link theory. 
{\it Surveys on surgery theory: surveys presented 
in honor of C.T.C. Wall Vol. 1, 345--364, Ann. of Math. Stud.,
145, Princeton Univ, 2000.}


\bibitem{Milnor} J. Milnor:
Lectures on the $h$-cobordism theorem 
{\it Princeton University Press} 1965


\bibitem{MilnorStasheff} J. W. Milnor and J. D. Stasheff:
 Characteristic classes. 
 {\it Annals of Mathematics Studies, No. 76. 
 Princeton University Press} 1974. 


\bibitem{Ogasa} E. Ogasa:  
(i) math.GT/0004008,{\it University of Tokyo Preprint series UTMS 97-35};
(ii){\it Mathematical Research Letters, 5 (1998), 577-582}, UTMS 95-50; 
(iii){\it Journal of knot theory and its ramifications, 10 (2001), 
121--132 } UTMS 97-34, math.GT/0003088; 
(iv){\it Mathematical Proceedings of Cambridge Philosophical Society 
126, 1999, 511-519}; 
(v)UTMS 97-63; 
(vi)math.GT/0011163,  UTMS00-65; 
(vii)math.GT/0004007, UTMS 00-22; 
(viii)hep-th/0311136. 


\bibitem{PeskinSchroeder}  M. E. Peskin and D. V. Schroeder: An introduction to quantum field theory. {\it Addison-Wesley Publishing Company}, 1995. 


\bibitem{Polchinski} J. Polchinski:  String theory. Vol. 1, 2.
{\it Cambridge, UK: Univ. Pr. }1998. 


\bibitem{Rolfsen} D. Rolfsen: Knots and links
{\it Publish or Perish, Inc.} 1976. 


\bibitem{Schwarz} J. H. Schwarz: The power of $M$ theory. 
{\it Phys. Lett. B 367 (1996), no. 1-4, 97--103}. 


\bibitem{Smale} 
S. Smale: Generalized Poincare's conjecture in dimensions greater than four. 
{\it Ann. of Math.} (2) 74 1961 391--406.


\bibitem{Steenrod} N. Steenrod: 
The topology of fibre bundles. 
{\it Reprint of the 1957 edition. Princeton University Press 1999.}





\bibitem{Townsend} 
P. K. Townsend:  M-theory: a new paradigm for quantum gravity. 
 {\it Frontiers in quantum physics (Kuala Lumpur, 1997), 15--23, Springer, Singapore, 1998}.


\bibitem{Vafa}  C. Vafa: Evidence for $F$-theory. 
{\it Nuclear Phys. B 469 (1996), no. 3, 403--415}. 


\bibitem{WittenQFTJones} E. Witten: 
 Quantum field theory and the Jones polynomial, 
 {\it Comm. Math. Phys.}, 121 (1989), 351-399.


\bibitem{Witten} E. Witten: 
String theory dynamics in various dimensions. 
{\it Nuclear Phys. B 443 (1995), no. 1-2, 85--126}.


\end{thebibliography}
\end{document}